\documentclass[%
 preprint,
superscriptaddress,
 amsmath,amssymb,
 aip,
rmp,
]{revtex4-2}

\usepackage{graphicx}
\usepackage{dcolumn}
\usepackage{bm}
\usepackage{verbatim}
\usepackage{color,soul}
\usepackage{epstopdf}
\AppendGraphicsExtensions{.tif}

\begin{document}


\title{The structure of Ferroelectric BaBiO$_3$/BaTiO$_3$ Interfaces grown by Molecular Beam Epitaxy}
\author{Merve Baksi}
 \affiliation{Department of Physics, North Carolina State University, Raleigh, NC 27695, USA}

\author{Divine P. Kumah}
\email{divine.kumah@duke.edu}
  \affiliation{Department of Physics, Duke University, Durham, NC 27517, USA}
  \affiliation{Department of Physics, North Carolina State University, Raleigh, NC 27695, USA}
 

\date{\today}

\begin{abstract}
We investigate the lattice structure of heterostructures comprising of ferroelectric BaTiO$_3$ (BTO) thin films and BaBiO$_3$ (BBO), the insulating parent-compound of the high Tc superconductor. Motivated by theoretical predictions of exotic phenomena in BBO-based heterostructures including interfacial conductivity, superconductivity and topologically-protected states, we synthesize BTO/BBO heterostructures by molecular beam epitaxy and characterize their structural properties by in-situ reflection high energy electron diffraction and ex-situ high-resolution synchrotron X-ray diffraction, Raman spectroscopy and piezoforce microscopy. For heterostructures with 4 uc BBO layers, reciprocal space maps indicate strain relaxation relative to the SrTiO$_3$ substrate. We observe a strong tetragonal distortion for heterostructures with 2 unit cells BBO coherently strained to the BTO layers. Raman spectroscopy measurements indicate a suppression of the breathing mode distortion associated with the charge density wave in bulk BBO. The ferroelectric properties of the system are confirmed by piezoforce microscopy measurements. The coupling between the ferroelectric polarization and the electronic states in BBO may potentially serve as a starting point for tunable electrostatic doping to realize the novel predicted states in the atomically-thin BBO layers.
\end{abstract}

\maketitle

\section{Introduction}
\begin{figure*}[ht]
\centering
\includegraphics[width=0.9\textwidth]{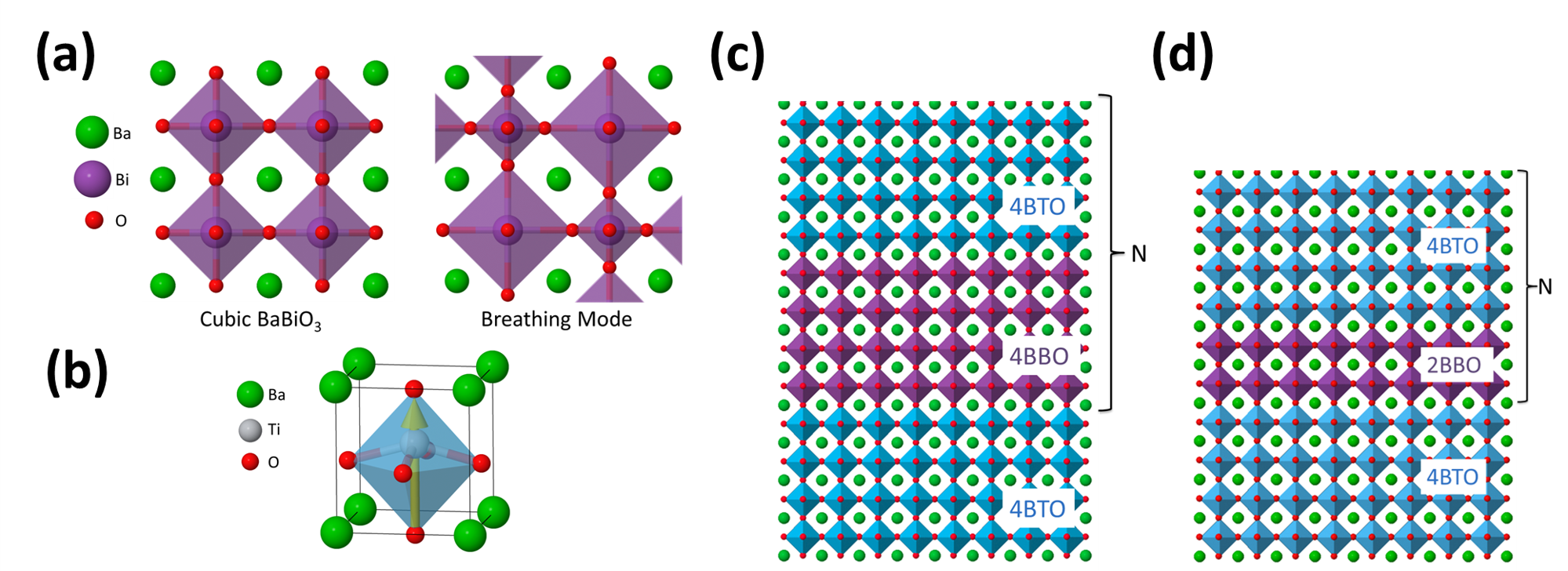}

\caption{(a)Crystal structure  of cubic and monoclinic BaBiO$_3$  and (b) ferroelectric tetragonal BaTiO$_3$. Schematic of heterostructures synthesized by molecular beam epitaxy with (c) 4 uc BTO and 4 uc BBO and (d) 4 uc BTO and 2 uc BTO.}
\label{fig:schematic}
\end{figure*}

Hole-doped perovskite BaBiO$_3$ (BBO) has been investigated for its superconducting properties. K and Pb doping leads to superconducting transition temperatures of 16 K (BaBi$_x$Pb$_{1-x}$O$_3$) and 32 K (Ba$_{1-x}$K$_{x}$BiO$_3$), respectively.\cite{bouwmeester2021babio3, sleight2015bismuthates, mattheiss1988superconductivity, cava1988superconductivity} The undoped parent compound has a monoclinic (\textit{I2/m}) structure with octahedral rotations and a breathing mode distortion as shown in Figure \ref{fig:schematic}(a) leading to an electronic gap and insulating behavior. The insulating state arises from a charge density wave (CDW) which has been proposed to arise from either bond disproportionation and/or charge ordering on the Bi sites with alternating Bi$^{3+}$ and Bi$^{5+}$.\cite{sleight2015bismuthates, foyevtsova2015hybridization, cox1976crystal, chaillout1985determination, shen1989photoemission} Hole-doping leads to a suppression of the CDW and metallicity, thus efforts to suppress the monoclinic distortion to induce an insulator-metal transition through epitaxial strain, confinement in atomically thin heterostructures and coupling to THz optical excitations have been experimentally and theoretically investigated.\cite{feng2022anti, kim2015suppression, zapf2019structural, bovzovic2020quest, bouwmeester2019stabilization}

Recently, theoretical predictions of novel electronic states have been reported in heterostructures based on atomically thin BBO. An interfacial two-dimensional electron gases at polar oxide/non-polar BBO (001) interfaces have been predicted where the polar oxide is the lattice-matched LaLuO$_3$.\cite{khazraie2020potential} Topologically protected states and superconductivity have been predicted in strained cubic BBO thin films\cite{yan2013large} and BBO/BaTiO$_3$(BTO) (111) bilayers.\cite{kim2018graphene} However, doping on the order of 0.5 electrons or holes per unit cell is required to place the Fermi level of the system at the predicted Dirac points which could be realized by chemical doping or electric gating. Currently, these predictions have not been verified experimentally, possibly due to questions about the thermodynamic stability of the proposed structures,\cite{malyi2020realization, malyi2020false} the large lattice mismatch that exists between BBO and commercially available substrates, and the proposed high doping level.\cite{bouwmeester2021babio3}

The structural and electronic coupling at epitaxial complex oxide interfaces provides a route to suppress the breathing mode distortion and simultaneously dope the BBO layers via electrostatic coupling and/or interfacial charge transfer. Hence, we investigate the structural properties of multilayer heterostructures comprising atomically-thin BBO layers and tetragonal ferroelectric BTO (bulk, \textit{P4mm }$a=b=3.992 \AA{}$ $c=4.036 \AA{}$)\cite{kwei1993structures} films grown by molecular beam epitaxy. Ideally, the lack of octahedral rotations in BTO will suppress the octahedral rotations in the BBO layers at the BTO/BBO heterointerface and the field effect arising from the ferroelectric polarization can potentially lead to non-volatile electrostatic doping of the BBO layers.\cite{vaz2021epitaxial} 

Surprisingly, while a large 10\% lattice mismatch exists between BBO (pseudocubic(\textit{pc}) a$_{pc}$=4.35 \AA{})\cite{cox1976crystal} and SrTiO$_3$ (STO)  (a=3.905 \AA{}), BBO films have been successfully grown on single-crystal STO substrates using pulsed laser deposition and molecular beam epitaxy \cite{zapf2018domain, gozar2007surface, yamamoto2004growth, kim2015suppression, jin2020atomic}. An interfacial layer is observed between the BBO and the substrate with the thickness of the transitional layer, $t_{int}$, on the order of 1-2 nm (2-4 BBO unit cells (ucs)). As the film thickness increases past $t_{int}$, BBO layers nucleate with bulk-like lattice parameters. Understanding the structure and composition of the interface layer is critical for understanding electron interactions with the substrate, and for designing strategies for synthesizing BBO films with thicknesses on the order of 1 unit cell in short-period superlattices to realize the exotic states predicted by first-principles theory.\cite{bovzovic2020quest} Transmission electron microscope measurements have been reported for BBO films on STO with varying structures.  Bouwmeester \textit{et. al.} propose an interface structure comprising of a rocksalt AO (where A may be Ba or Sr) bilayer on the STO substrate prior to the nucleation of the BBO layer. Strain relaxation is proposed to occur in the second AO layer.\cite{bouwmeester2019stabilization}   Jin \textit{et. al.} propose an interfacial structure comprising of a double BiO layer with a $\delta$-Bi$_2$O$_3$ (BO)-like phase with fluorite structure. \cite{jin2020atomic} However, the authors point out that the interface thickness and crystallinity is dependent on the Ga ion milling procedure used to prepare the samples for the TEM measurements.

In this work, we investigate the structural properties of \textit{m}  BTO/\textit{n} BBO heterostructures on (001)-oriented STO substrates by molecular beam epitaxy.  The BBO thickness, \textit{n}, is constrained to be on the order of the reconstructed interface layer between 2-4 unit cells (uc) with the BTO thickness, \textit{m}, is fixed at 4 ucs. Schematics of the structures studied are shown in Figure \ref{fig:schematic}(c) and (d) for n=4 and 2, respectively. The interface structure is determined using synchrotron X-ray diffraction which permits a non-destructive determination of the atomic structure and composition of the buried interfaces. For a single 4 uc  BTO buffer/ 4 uc BBO /4 uc BTO trilayer on STO ([4/4/4]x1), Bi-Ti intermixing between the BBO layer and the capping layer leads to strain relaxation. As the number of BTO/BBO bilayer increases in [4 BTO/4 BBO]x5 ([4/4]x5) superlattices (SLs), both BBO and BTO relax to their bulk values as evidenced by in-situ reflection high energy diffraction (RHEED) measurements and X-ray reciprocal space maps. On reducing the BBO thickness to 2 uc in [4 BTO/ 2 BBO]x5 ([4/2]x5), two BBO domains are observed. The first domain is fully relaxed, while the second domain is compressively strained to the BTO layer with a strong tetragonal distortion. Piezoforce microscopy (PFM) measurements indicate that the BTO layers retain their ferroelectric properties. While Raman spectroscopy measurements indicate a suppression of the breathing mode associated with the CDW and the electronic gap in undoped BBO, the films remain insulating.

\section{Experimental Details}
\subsection{Synthesis}
BTO/BBO/BTO trilayers and [4/4]x5 and [4/2]x5 superlattices were grown by molecular beam epitaxy on (001)-oriented single-crystal SrTiO$_3$ substrates. The BBO thickness was varied from 2-4 unit cells and the BTO thickness was fixed at 4 unit cells.  High-purity Bi, Ba, and Ti were evaporated from effusion cells. The fluxes were calibrated before the film deposition using a quartz crystal monitor. To account for Bi desorption at the growth temperature, the Bi:Ba ratio was fixed at 2:1 based on optimizing the crystallinity and stoichiometry of 40 uc thick BBO films. The BBO and BTO films were deposited at 650 C in 6.5$\times$ 10$^{-6}$ Torr Oxygen plasma. After growth, the films were cooled to room temperature in the oxygen plasma. The SLs were capped with 4 uc BTO. The film crystallinity during the growth was monitored by reflection high energy electron diffraction.

\subsection{Structural Measurements}
Crystal truncation rods (CTRs)\cite{robinson1992surface} and reciprocal space maps (RSMs) were measured at the 33ID beamline at the Advanced Photon Source at Argonne National Laboratory to determine the structural properties of the films. The measurements were performed at room temperature with an incident photon energy of 13.5 KeV. The CTRs were fit using the GenX genetic fitting algorithm \cite{bjorck2007genx, kumah2011correlating, Koohfar2019Confinement} to determine the structures of the epitaxial fractions of the films.
Raman spectroscopy measurements were performed at room temperature using a Horiba Confocal Raman Microscope with a 532 nm laser.

\begin{figure}[h]

\includegraphics[width=6in]{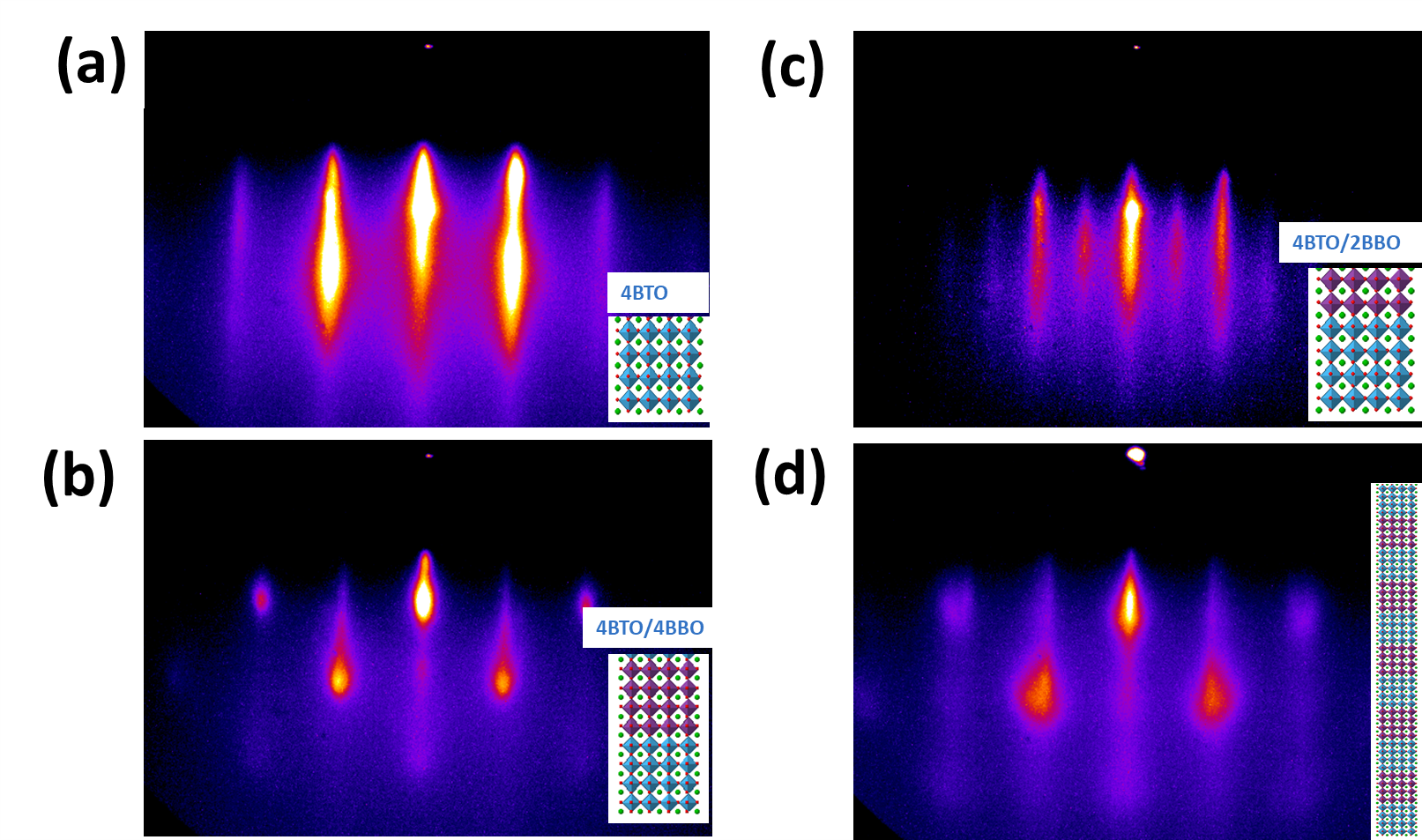}

\caption{ RHEED pattern of [4 BaTiO$_{3}$/4 BaBiO$_{3}$]$_5$/4 BaTiO$_{3}$ heterostructure on (001)-oriented STiO$_3$  after growth of (a) the first 4 ucs of  BaTiO$_{3}$ (b)the first 2 ucs of  BaBiO$_{3}$ (c) 4 layer of BaBiO$_{3}$. (d) Final RHEED image at the end of the growth of the superlattice and the 4 uc BaTiO$_{3}$ capping layer. }
\label{fig:bbo64_rheed}
\end{figure}

\section{Results and Discussion}
\subsection{RHEED Results}

RHEED images taken during the growth of [4/4]x5  superlattice are shown in Figure  \ref{fig:bbo64_rheed}. Figure \ref{fig:bbo64_rheed}(a) show the RHEED images after the growth of the first 4 uc of BTO on STO.  The film crystallinity is evident by the sharpness of the RHEED streaks. During the deposition of BBO on the BTO buffer layer, the RHEED pattern evolves from sharp streaks after 2 unit cells (Figure \ref{fig:bbo64_rheed}(b)) to diffraction spots superimposed on the RHEED streaks ((Figure \ref{fig:bbo64_rheed}(c))) after 4 ucs. The transition is indicative of roughening of the surface as the thickness is increased. The final RHEED image after completion of the growth of the[4/4]x5 SL is shown in Figure \ref{fig:bbo64_rheed}(d). The in-plane lattice constant determined from the spacing of the RHEED diffraction pattern alternates between $\sim$ 3.98 $\pm0.02$\AA{} during the deposition of the BTO layers to $\sim$ 4.35 $\pm0.01$ \AA{} for the BBO layers, indicating structural relaxation of the BBO and BTO layers in the superlattice structure relative to the STO substrate. 

   \begin{figure*}[th]
\includegraphics[width=6.5in]{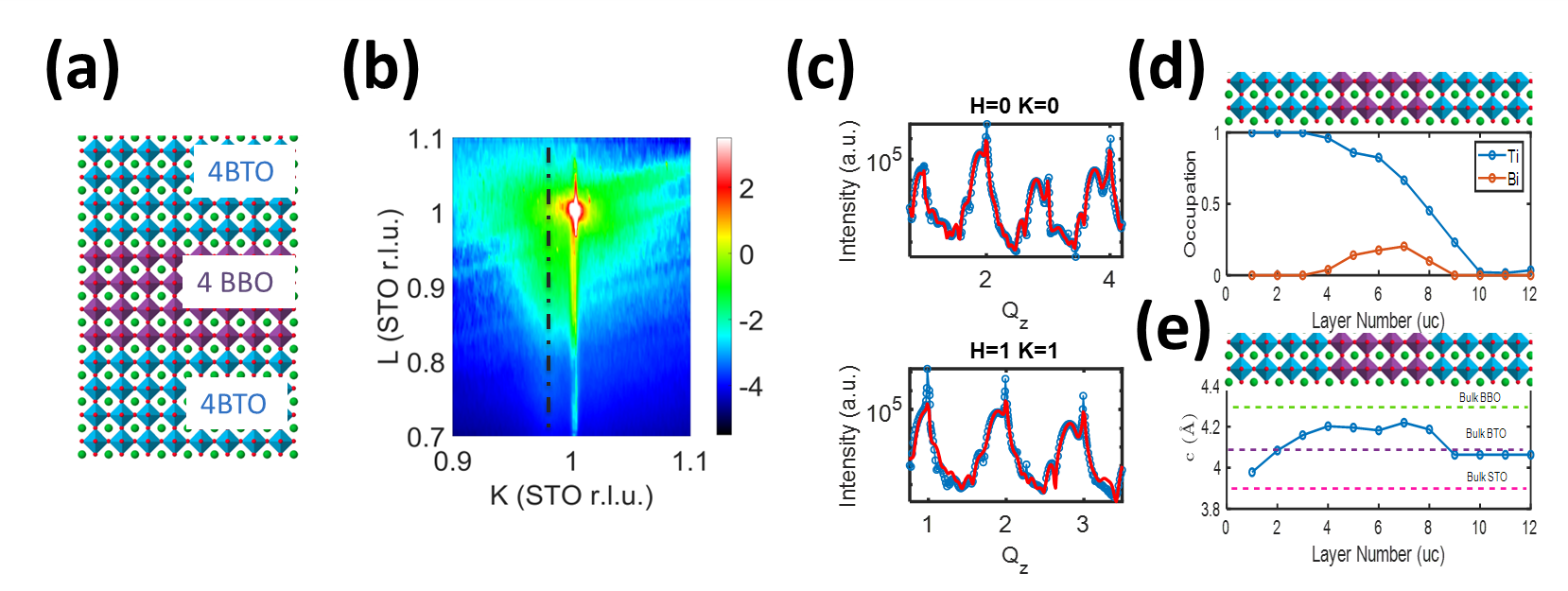}
\caption{(a)Schematic of 4 BTO/4 BBO/ 4 BTO trilayer structure grown on (001)-oriented SrTiO$_3$ (b) Reciprocal space map around the STO (0 1 1) Bragg peak showing the crystal truncation rod along K=1 and the relaxed diffraction peak (indicated by dashed line) along K=0.98. (c) Representative CTRs measured (blue circles) along the STO ( 0 0 \textit{L}) and (1 1 \textit{L}) directions and calculated intensities (red) for the best fit model of the coherently strained fraction of the trilayer. (d) Measured composition profile for the coherently strained component of the film and (e) layer-resolved out-of-plane lattice parameters determined from the fits in (c). The dashed lines indicate the bulk lattice parameters of STO, BTO and BBO.}
\label{fig:TL444}
\end{figure*}

\subsection{Analysis of [4 BTO/4 BBO/4 BTO] Trilayer}

Synchrotron X-ray CTRs were measured along the 00L, 10L, 11L, and 20L directions for the nominal 4 BBO/4 BTO/ 4 BTO trilayer sample. A schematic of the sample is shown in \ref{fig:TL444}(a). The CTRs with in-plane lattice vectors fixed to reciprocal lattice units (r.l.u.) defined by the bulk STO substrate (1 r.l.u.=$1/3.905 \AA{}$) measure the \textit{crystalline fraction of the film} coherently strained to the substrate. Thus, fractions of the film that are relaxed and/or incoherently strained do not contribute to the intensities along the CTRs. The reciprocal space map around the STO (1,0,1) Bragg peak is shown in Figure \ref{fig:TL444}(b). The intensities corresponding to the coherently strained fraction of the film lie along H=1 STO rlu. A broad diffuse peak corresponding to the relaxed fraction is observed centered along H=0.98 STO rlu corresponding to a 3.98 \AA{} in-plane lattice. The in-plane lattice constant is consistent with the RHEED-determined lattice constant of the surface layer, thus, we conclude that the fractions of the surface layers are relaxed, relative to the STO substrate. 

Figure \ref{fig:TL444}(c) shows representative CTRs measured for the  4 BBO/4 BTO/ 4 BTO sample and intensities calculated for the best-fit structural model. The initial model for fitting the CTR data comprised of fully strained BTO and BBO layers with the nominal thicknesses and stoichiometry does not match the periodicity of the measured intensities. However, allowing the \textit{B}-site (Ti, Bi) occupation to vary and non-unity layer occupations of the surface layers yields a good fit to the measured data. 
 Figure \ref{fig:TL444}(d) and (e) shows the layer-resolved Ti and Bi fractional occupations and the lattice spacings, respectively, determined by the differences in the z-positions of the A-site (Ba) along the growth direction. The coherently strained fraction in each layer decreases from layer 5 to layer 10 where BBO and the top BTO layer should be nominally located. The  Bi content at the \textit{B}-site is less than 20\%, indicating possible incorporation of Ti from the BTO cap layer in the BBO film. The average composition in the nominal BBO layer (layer 5-8)  is Ba(Ti$_{0.8}$Bi$_{0.2}$)O$_{3}$. The lattice constant increases in the 1st BTO layer from 4 \AA{} to 4.2 \AA{} in the nominal Ba(Ti$_{0.8}$Bi$_{0.2}$)O$_{3}$ layer.

\subsection{Structure of [4 BTO/4 BBO]N  superlattice}

 \begin{figure}[t]
\includegraphics[width=6in]
{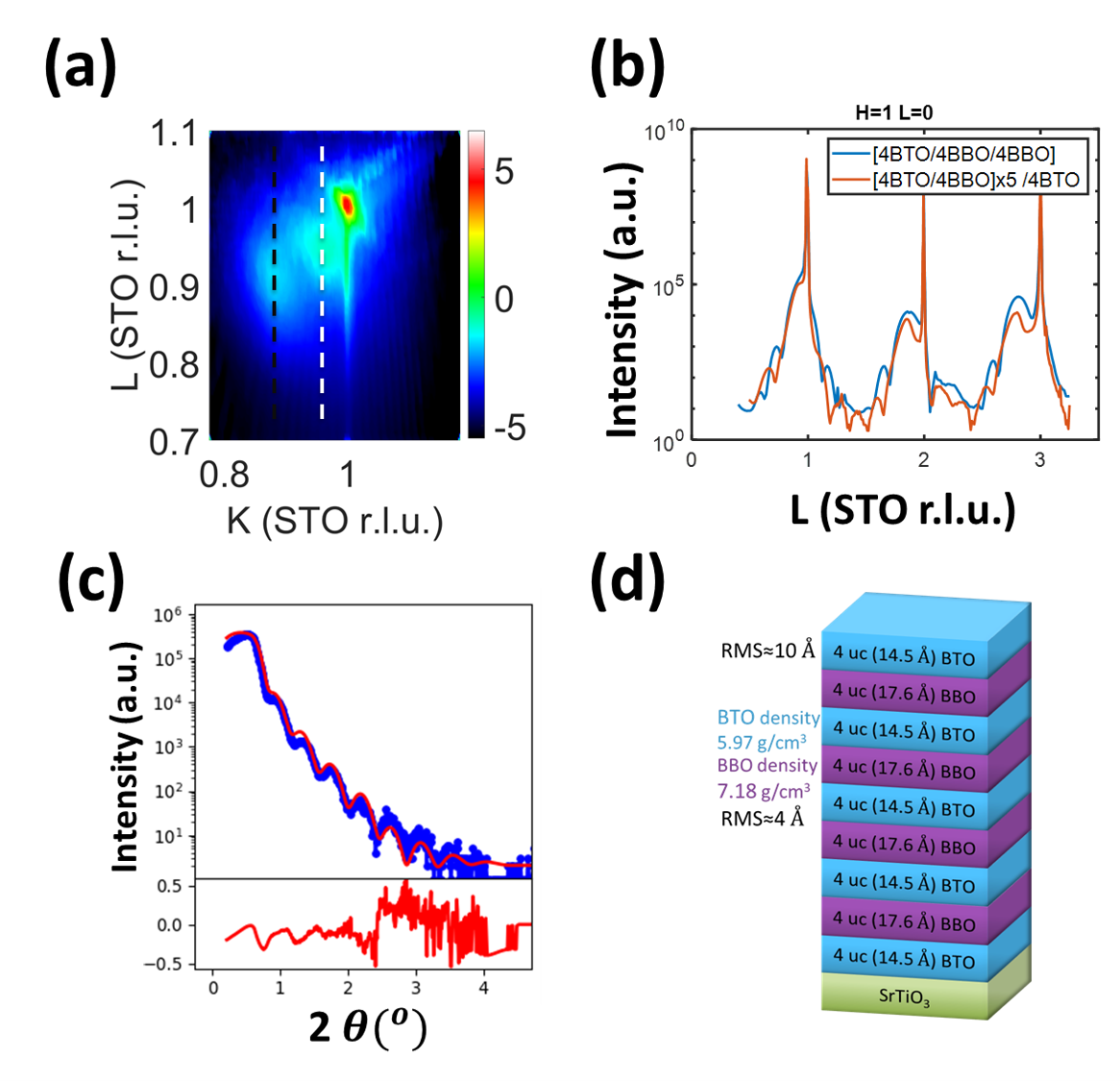}
\caption{(a) Reciprocal map around the STO (0 1 1) Bragg peak for a  [4BTO/4BBO]$_5$ superlattice The relaxed BBO and BTO diffraction peaks are indicated by the vertical black and white dashed lines, respectively. (b) Comparison of the (1 0 $L$) crystal truncation rods for the [4/4/4] trilayer and [4/4]$_5$ superlattice (c) Measured (blue circles) and calculated (red line) X-ray reflectivity intensities and (d) converged model structure for the [4BTO/4BBO]$_5$ superlattice. }
\label{fig:SL44}
\end{figure}

Next, we investigate the structure of a [4/4]x5 SL capped with 4 uc BTO.  The increase by a factor of 5 of the BTO/BBO bilayer thickness should result in a decrease in the film peak width along the L direction and an increase in the film Bragg peak intensity for the off-specular CTRs. However, we find that the CTRs along the off-specular directions are identical to the trilayer sample, as shown in Figure \ref{fig:SL44}(a) indicating that only the initial layers are coherently strained to the STO substrate.  The reciprocal space map for the sample is shown in  \ref{fig:SL44}(b). The narrow peak centered at (1,0,0.98) corresponds to intensities from the coherently strained fraction of the film. The STO substrate Bragg peak is located at (1,0,1). Two broad film peaks are located at (0.890,0,0.901)  corresponding to relaxed BBO with a$_{pc}$=4.38 \AA{} and c$_{pc}$=4.34 \AA{} and (0.960,0,0.955) corresponding to relaxed tetragonal BTO with a=4.06 \AA{} and c=4.09 \AA{}. The expanded volume of the BTO layer relative to bulk, suggests Bi interdiffusion into the BTO layer.

X-ray reflectivity measurements validate the nominal layer sequence. Figure \ref{fig:TL444} (c) shows the measured reflectivity for the [4/4]x5  and the calculated reflectivity for the best-fit model comprising of the nominal layer sequence. The fitted thicknesses for the BBO and BTO layers are 16.5 \AA{} and 17.6 \AA{}, in agreement with the nominal values. The layer roughness determined from the fits are on the order of 1-2 unit cells. Thus, we conclude, based on the X-ray and RHEED data, that the BTO/BBO lattice grows with the individual layers relaxed relative to the STO substrate.

\subsection{Structural analysis of [4 BTO/2 BBO]5 superlattice }
\begin{figure}[ht]
\centering
\includegraphics[width=2in]
{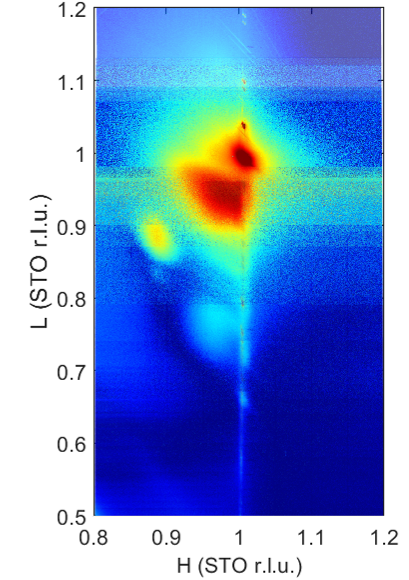 }
\caption{(a) Reciprocal space map around (101) Bragg peak for [4 BTO/ 2 BBO/ 4 BTO]x5 superlattice on STO.}
\label{fig:SL42}
\end{figure}

To understand the effect of the BBO thickness on the the strain state of the system, we investigate the structural properties of a [4/2]x5 SL. Figure \ref{fig:SL42}(a) shows the reciprocal space map for the [4/2]x5 SL sample around the substrate (1, 0, 1) Bragg peak. The BTO peak is located at (0.981, 0 0.944) STO rlu and is relaxed relative to the STO substrate. The lattice constants for the BTO layers are a=3.982 \AA{} and  c=4.135 \AA{}. A 2nd  peak at (0.896,0,0.903) corresponds to relaxed BBO with a=4.358 $\pm$0.01 \AA{} and c = 4.325 $\pm$0.01 \AA{}. The lattice parameter is comparable to pseudocubic lattice constant of the BBO bulk monoclinic phase,$a_{pc}=$4.35 \AA{}.\cite{pei1990structural} The slight increase may be attributed to Bi vacancies\cite{zapf2019structural} or cubic BBO\cite{kim2015suppression}. 
In contrast to the [4/4]x5 SL, a weak diffraction peak is observed for the [4/2]x5 SL at (0.957, 0, 0.765) STO rlu with lattice constants a= 4.0386 \AA{} and c= 5.1121 \AA{} . This peak is associated with BBO compressively strained to BTO.   The calculated volume of the strained BBO is 83.374 $\AA{}^3$ in close agreement with the bulk value of 82.59 $\AA{}^3$.\cite{pei1990structural} 

\begin{figure}[h]
\centering
\includegraphics[width=0.45\textwidth]{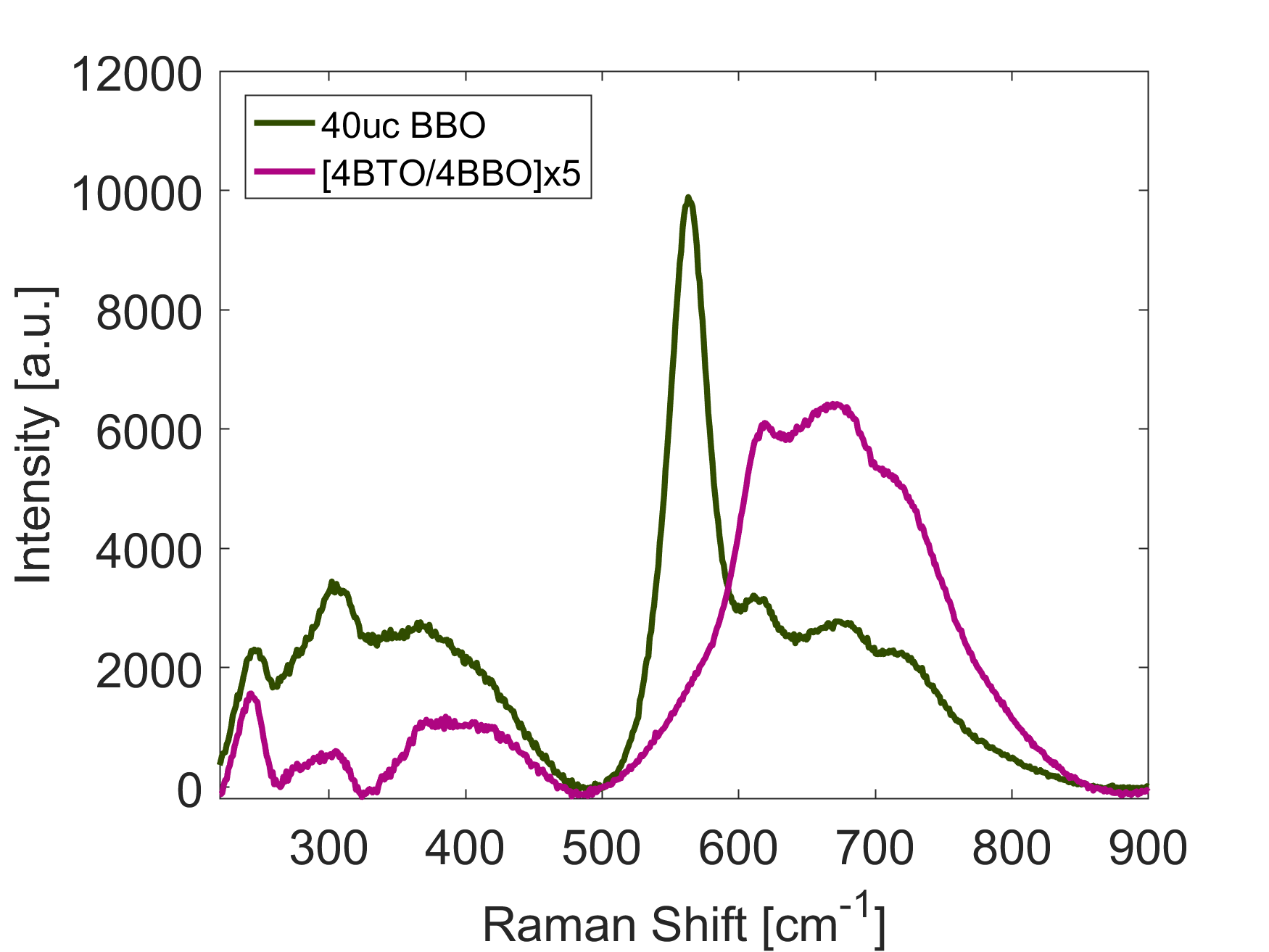}
\hspace{10pt}
\caption{Raman spectra of [4 BTO/4 BBO]x5 /4 BTO heterostructure and 40 uc of BaBiO$_3$ grown on (001)-oriented STO. }
\label{fig:Raman}
\end{figure}

The X-ray diffraction measurements confirm the growth of relaxed crystalline BBO/BTO bilayers with 2-4 uc thick BBO layers thicknesses. To determine if the structural breathing distortion associated with the CDW is suppressed in the ultrathin BBO layers, we compare Raman spectroscopy measurements for a "bulk" 40 uc BBO film and the [4/4]x5 SL in Figure \ref{fig:Raman}. An A1g Raman-active mode is observed for bulk BBO at 570 cm$^{-1}$, which is attributed to the breathing phonon mode.\cite{lederle1994raman, tajima1992raman} The mode is observed for the 40 uc BBO film on STO as shown in Figure \ref{fig:Raman}. Additional peaks below and above 570 cm$^{-1}$ are related to the second-order Raman-active modes in the bulk cubic STO substrate\cite{Yuzyuk2012, Krain2020}. In contrast to the 40 uc BBO sample, the A1g Raman active mode is strongly suppressed in the [4/4]x5 SL as shown in Figure \ref{fig:Raman}. A similar suppression has been reported in hole-doped cubic superconducting BaK$_x$Bi$_{1-x}$O$_3$\cite{tajima1992raman} where the CDW is suppressed and metallicity emerges. 

The absence of the A1g Raman peak in BBO films grown on STO with thicknesses below a critical thickness of 11 unit cells was previously attributed to a dimensionality induced monoclinic to cubic ($Pm\bar{3}m$) transition and the suppression of the CDW.\cite{kim2015suppression} However, the formation of an interfacial layer between BBO and STO\cite{jin2020atomic} and Bi off-stoichiometry may explain the absence.\cite{zapf2019structural} In our current work, we observe a clear structural peak related to stoichiometric BBO in the [4/4]x5 (Figure \ref{fig:SL44}) superlattices without the observed A1g peak. However, the slightly larger lattice constants and lattice volume compared to bulk BBO may be  associated with  Bi-deficient Ba$_{1-x}$Bi$_x$O$_y$ films.\cite{zapf2019structural} \textit{Our superlattices are insulating and exhibit a strong ferroelectric response.}

\subsection{PFM measurements}
\begin{figure}[ht]
\includegraphics[width=6in]{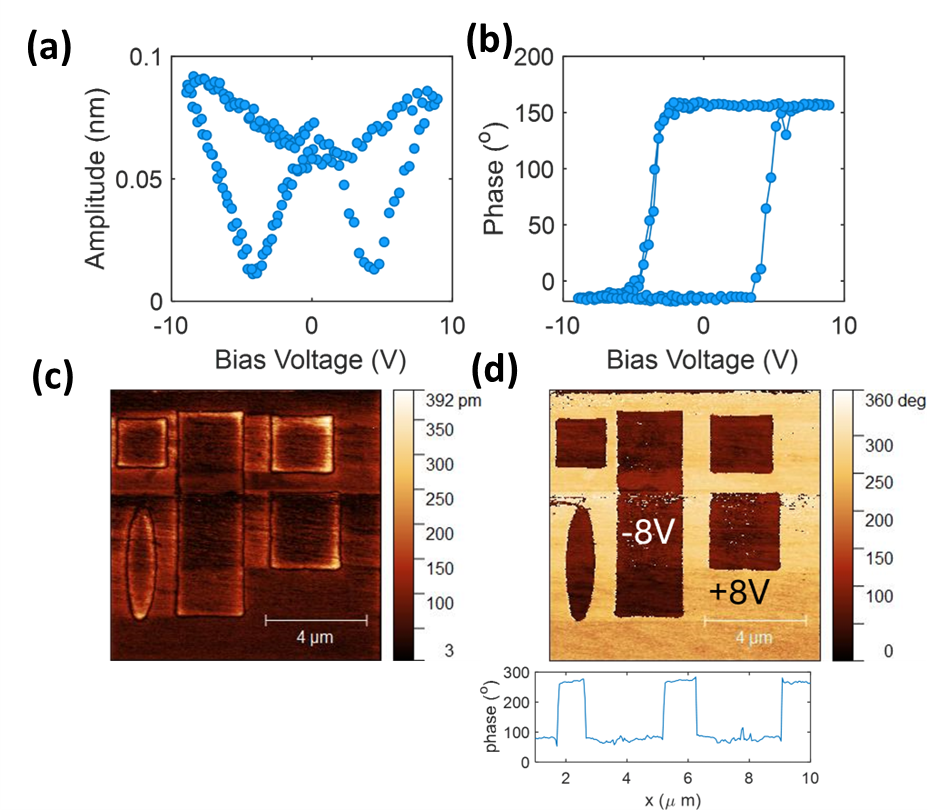}
\caption{ Piezoforce microscopy measurements of [4 BTO/ 4 BBO]x5 on STO (a) PFM amplitude and (b) Phase as a function of applied bias. (c)Amplitude and (d) phase contrast measured after writing domains with a +/-8 V bias. }
\label{fig:PFM}
\end{figure}

Finally, we confirm the ferroelectric properties of the samples by piezoforce microscopy. Figure \ref{fig:PFM} shows the amplitude and phase contrast measured after writing with $\pm$8 V bias and reading with  0.5 V$_{AC}$ driving voltage. Up and down ferroelectric domains are clearly visible. The PFM loops show clear ferroelectric hysteresis and ferroelectric switching at 4 V indicating that the BTO retains its ferroelectric properties in the BTO/BBO heterostructures. The ferroelectric field effect potentially provides an effective route to accumulate excess holes and electrons in the interfacial BBO layers without 
 disorder related to chemical doping. \cite{hoffman2010ferroelectric} In the symmetric [BTO/BBO/BTO] heterostructures, the separation of the BBO hole-rich and electron-rich interfaces must be above a critical thickness to induce an insulator-to-metal transition. Theoretical investigations of electron and hole gases formed at polar LaLuO$_3$/SrBiO$_3$ interfaces estimate a critical thickness on the order of 4 unit cells.\cite{khazraie2020potential} Thus, tuning of the BBO thickness and/or asymmetric ferroelectric/BBO heterostructures may be necessary to induce two-dimensional electron and hole gases in the ultrathin BBO layers.

\section{Conclusion}
In conclusion, we have investigated the structural properties of BTO/[m BBO/ 4 BTO]xN heterostructures grown by molecular beam epitaxy on (001)-oriented STO substrates. The nominal lattice mismatch between the BBO and STO substrate is 12\%. Relaxation is observed after deposition of the 1st bilayer by in-situ RHEED and ex-situ reciprocal space maps. While the stoichiometry is preserved for 2 uc BBO on the 4 uc BTO buffer on STO, Bi/Ti intermixing leads to coherent epitaxy in the 1st bilayer when the BBO thickness is increased to 4 unit cells.  For [2 BBO/ 4 BTO]x5 SLs, the BBO is compressively strained to the BTO leading to tetragonal BBO layers with a c/a ratio of 1.25. For [4 BBO / 4 BTO] x5 SL with increased thicknesses of the BBO, the increase in strain energy leads to relaxation of the BBO layers to the monoclinic bulk phase with a$_{pc}$ = c$_{pc}$ = 4.38 \AA{}. 
The breathing mode distortions are suppressed in the relaxed confined BBO layers for thicknesses up to 6 unit cells.
While a tetragonal distortion is observed from the [2BBO/4 BTO] heterostructure, no metallicity is currently observed and doping strategies via the field effect and/or chemical substitution may be required to induce metallicity and/or the proposed topologically protected phase in BBO heterostructures.\cite{bouwmeester2021babio3}

\section{Acknowledgements}
Use of the Advanced Photon Source was supported by the US Department of Energy, Office of Science, Office of Basic Energy Sciences, under Contract No. DE-AC02-06CH11357. This work was performed in part at the Analytical Instrumentation Facility (AIF) at North Carolina State University, which is supported by the State of North Carolina and the National Science Foundation (award number ECCS-2025064). This work made use of instrumentation at AIF acquired with support from the National Science Foundation (DMR-1726294). The AIF is a member of the North Carolina Research Triangle Nanotechnology Network (RTNN), a site in the National Nanotechnology Coordinated Infrastructure (NNCI).

\section{Data Availability Statement}
The data that support the findings of this study are available from the corresponding author upon reasonable request

\end{document}